\documentclass[prl,twocolumn,showpacs,groupedaddress,notitlepage,superscriptaddress,floatfix,nofootinbib]{revtex4-2}
\usepackage{amsmath,amsthm,amssymb,amsfonts,enumitem,float,graphicx,physics,hyperref}
\usepackage{color}
\usepackage[dvipsnames]{xcolor}
\usepackage[normalem]{ulem}
\usepackage{subfigure}
\usepackage{csquotes}

\hypersetup{
    colorlinks=true,       
    linkcolor=cyan,          
    citecolor=magenta,        
    filecolor=magenta,      
    urlcolor=cyan,           
    runcolor=cyan
}

\usepackage{amsthm}
\usepackage{physics}
\usepackage[capitalise]{cleveref}

\usepackage{xcolor}

\theoremstyle{definition}

\DeclareMathOperator{\TTE}{\text{TT}\varepsilon}

\newcommand{\al}{\alpha}

\newcommand{\q}[1]{\enquote{#1}}

\newcommand{\ignore}[1]{}
\newcommand{\beq}{\begin{equation}}
    \newcommand{\eeq}{\end{equation}}
\newcommand{\bes} {\begin{subequations}}
    \newcommand{\ees} {\end{subequations}}

\makeatother

\begin{document}
    
    \title{Scaling Advantage in Approximate Optimization with Quantum Annealing}
    
    \author{Humberto Munoz Bauza}

    \affiliation{Department of Physics and Astronomy, University of Southern California,
        Los Angeles, California 90089, USA}
    
    \affiliation{Center for Quantum Information Science \& Technology, University
        of Southern California, Los Angeles, California 90089, USA}
        
    \author{Daniel Lidar}
    
    \affiliation{Department of Physics and Astronomy, University of Southern California,
        Los Angeles, California 90089, USA}
    
    \affiliation{Center for Quantum Information Science \& Technology, University
        of Southern California, Los Angeles, California 90089, USA}
    
    \affiliation{Department of Electrical Engineering, University of Southern California,
        Los Angeles, California 90089, USA}
    
    \affiliation{Department of Chemistry, University of Southern California, Los Angeles,
        California 90089, USA}

\begin{abstract}
Quantum annealing is a heuristic optimization algorithm that exploits quantum evolution to approximately find lowest energy states~\cite{kadowakiQuantum98,Hauke:2019aa}. 
Quantum annealers have scaled up in recent years to tackle increasingly larger and more highly connected discrete optimization and quantum simulation problems~\cite{johnsonQuantum11,boixoEvidence14,kingObservation18,harrisPhase18,Weinberg:2020aa}. 
Nevertheless, despite numerous attempts, a computational quantum advantage in exact optimization using quantum annealing hardware has so far remained elusive~\cite{speedup,katzgraberSeeking15,Venturelli:2014nx,PhysRevX.6.031015,mandraPitfalls17,albashDemonstration18,mandraDeceptive18,kowalsky20213regular,ebadiQuantum22}. 
Here, we present evidence for a quantum annealing scaling advantage in approximate optimization. The advantage is relative to the top classical heuristic algorithm: parallel tempering with isoenergetic cluster moves (PT-ICM)~\cite{zhuEfficient15}. The setting is a family of 2D spin-glass problems with high-precision spin-spin interactions.
To achieve this advantage, we implement quantum annealing correction (QAC)~\cite{pudenzErrorcorrected14}: an embedding of a bit-flip error-correcting code with energy penalties that leverages the properties of the D-Wave Advantage quantum annealer to yield over $1,300$ error-suppressed logical qubits on a degree-$5$ interaction graph. 
We generate random spin-glass instances on this graph and benchmark their time-to-epsilon, a generalization of the time-to-solution metric~\cite{speedup} for low-energy states.
We demonstrate that with QAC, quantum annealing exhibits a scaling advantage over PT-ICM at sampling low energy states with an optimality gap~\cite{MIPgap} of at least $1.0\%$. 
This amounts to the first demonstration of an algorithmic quantum speedup in approximate optimization.
\end{abstract}

\maketitle

The pursuit of a quantum speedup using quantum processors is a major theme in modern physics and computer science. Two of the leading application areas are discrete optimization and quantum simulation.
In the 
latter context, impressive recent quantum annealing (QA) advances have been made for fast, coherent anneals lasting on the order of the single superconducting flux qubit coherence time~\cite{kingCoherent22,kingQuantum22}. While this diabatic approach is considered promIsing~\cite{crossonProspects20}, 
it cannot be expected to scale up without the introduction of error correction or suppression, as decoherence and control errors pose insurmountable challenges to Hamiltonian quantum computation models~\cite{childs_robustness_2001,amin_decoherence_2009,albashDecoherence15,zhuBestcase16,Albash:2017ab,Albash:2019ab},
just as they do for gate-model quantum computers. In the absence of a fault-tolerance threshold theorem~\cite{Campbell:2017aa}
for QA, a variety of Hamiltonian error suppression techniques have been proposed and analyzed as ways to reduce the error rates of this computational model and the closely related model of adiabatic quantum computation~\cite{jordanErrorcorrecting06,PhysRevLett.100.160506,Young:13,Ganti:13,bookatzError15,Jiang:2015kx,marvianError17a},
providing tools towards scalability. 

However, despite these theoretical advances, there are currently practical limitations to the types and locality of programmable interactions in the Hamiltonians of quantum annealing hardware, even as new devices have started to emerge~\cite{Glaetzle:2017aa,Hamerly:2019aa,Tennant:2022uw,ebadiQuantum22}.
To address these limitations, quantum annealing correction (QAC)~\cite{pudenzErrorcorrected14} was developed as a realizable error suppression method for quantum annealing, targeting the available and restricted set of control operations in quantum annealers. QAC has been demonstrated to enhance the success probability of quantum annealing and mitigate the analog control problem~~\cite{pudenzErrorcorrected14,vinciNested16,pearsonAnalog19}.
The QAC method is based on a repetition-code encoding of the Hamiltonian and does not fully realize a Hamiltonian stabilizer code. Despite this, it has been shown theoretically to increase the energy gap of the encoded Hamiltonian and reduce tunneling barriers, thus softening the onset of the associated critical dynamics as well as lowering the effective temperature~\cite{Matsuura:2016aa}.
     
Here, departing from the traditional paradigm of using QA for exact optimization, we demonstrate---by incorporating QAC---the first genuine scaling advantage in \emph{approximate} optimization (low-energy sampling) using a quantum annealer. Even approximate optimization can be computationally hard unless $\mathrm{P=NP}$~\cite{trevisanInapproximability04,aroraComputational09}, so we do not expect the advantage we exhibit to amount to more than a polynomial speedup. However, 
whereas the scaling advantages reported in previous work were relative to simulated annealing~\cite{albashDemonstration18,ebadiQuantum22}, the advantage we find here is over the best currently available general heuristic classical optimization method: 
parallel tempering with isoenergetic cluster moves (PT-ICM)~\cite{zhuEfficient15}.
This result is enabled by implementing QAC on the D-Wave Advantage quantum annealer for the Sidon-set spin glass instance class~\cite{katzgraberSeeking15}, embedded on the logical graph formed after the encoding step.
The advantage of quantum annealing over PT-ICM is diminished without QAC, thus highlighting the crucial role of quantum error suppression.\\  
    
    \begin{figure*}[t]
    \includegraphics[width=\linewidth]{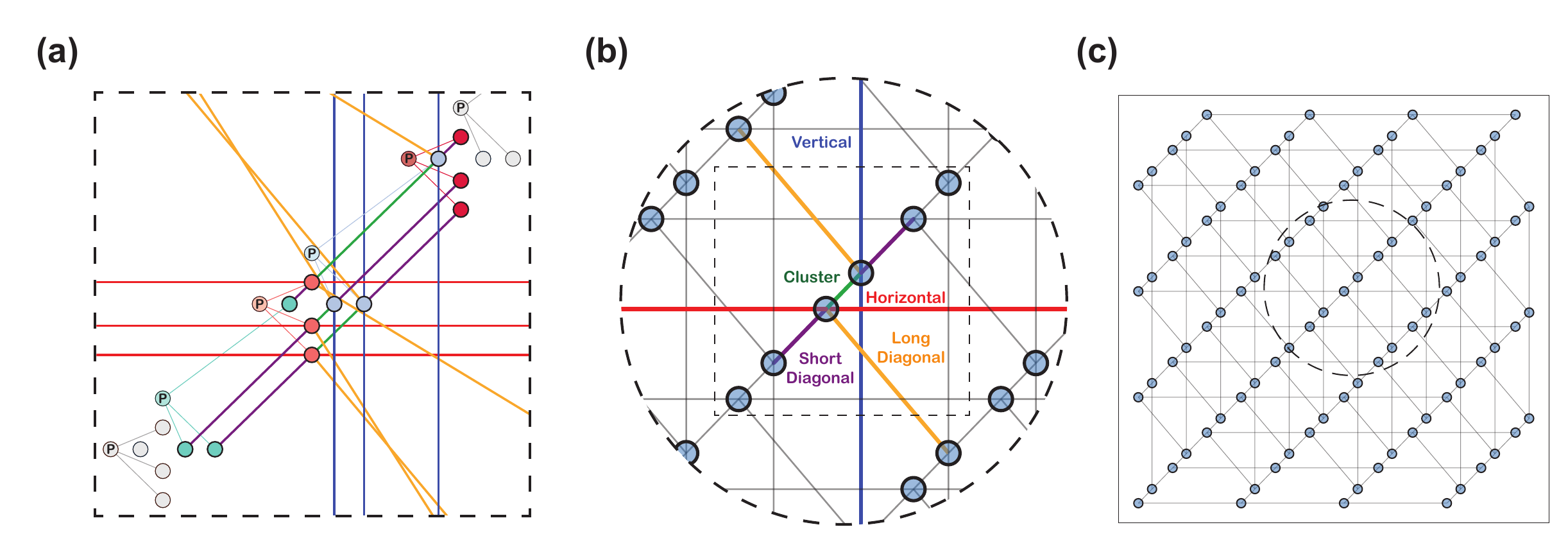}
    \caption{
        Embedding of QAC on the Pegasus hardware graph.         
         (a) Close-up of the embedding of three clusters of two coupled logical qubits each. Physical qubits are denoted by circles and couplings by lines. Penalty qubits are marked by a P and the three penalty couplings (thin lines) to their corresponding physical data qubits. Thick lines indicate the physical couplings between the corresponding physical qubits of different logical qubits. Only a subset of all possible couplings are shown.
        (b) A zoomed-out view from (a) showing the logical qubits (circles) and the logical couplings induced by the QAC embedding. Thick lines indicate the logical couplings shown in (a), whose type is colored by direction: horizontal/vertical/diagonal (long or short).
        (c) A zoomed-out view from (b) showing a $4\times 4$ induced logical graph of QAC. 
        The logical graph is equivalent to a honeycomb graph with additional non-planar bonds. The induced logical graph of the D-Wave Advantage 4.1 QPU has a maximum of $1322$  logical qubits; the largest available side length is $L=15$.
    }
    \label{fig:pegasus-qac}
    \end{figure*}

\noindent\textbf{\large  Quantum annealing}\\
The D-Wave quantum processing unit (QPU) uses superconducting flux qubits to implement the transverse field Ising Hamiltonian
    \begin{equation}
    H(s) =  -A(s) \sum_{i\in \mathcal{V}} \sigma^x_i + B(s) H_z ,
    \end{equation}
    where $\mathcal{V}$ is the vertex set of the hardware graph of the QPU, $i$ is the qubit index, $\sigma^x_i$ are Pauli matrices, and $A(s)$ and $B(s)$ are the annealing schedules, respectively decreasing to $0$ and increasing from $0$ with $s:0\to 1$. $H_z$ is the Ising problem Hamiltonian:
    \begin{equation}
        H_z = \sum_{i\in \mathcal{V}} h_i \sigma_i^z + \sum_{\{i,j\}\in \mathcal{E}} J_{ij} \sigma_i^z \sigma_j^z ,
    \end{equation}
    where $h_i$ and $J_{ij}$ are programmable local fields and couplings, respectively, and $\mathcal{E}$ is the edge set of the hardware graph.
    Many NP-complete and NP-hard problems can be mapped to $H_z$~\cite{2013arXiv1302.5843L} by minor-embedding onto the hardware graph. 
    We performed QA experiments on the D-Wave Advantage 4.1 QPU accessed through the D-Wave Leap cloud interface,
featuring the Pegasus hardware graph~\cite{boothbyNextGeneration20}.\\
    
\noindent\textbf{\large Quantum annealing correction}\\
 We implement the $[[3,1,3]]_1$ QAC encoding introduced in Ref.~\cite{pudenzErrorcorrected14}, which encodes a logical qubit into three physical \q{data qubits,} each of which is coupled to the same additional \q{energy penalty qubit} with a fixed coupling strength $J_p$; the logical qubit is decoded via a majority vote on the data qubits.
    The logical subgraph induced by the QAC encoding on the Pegasus graph has a bulk degree of $5$ and admits native loops of length $5$. 
    \cref{fig:pegasus-qac} illustrates the encoding and the induced logical graph.
    All previous QAC results
    were obtained using the \q{Chimera} hardware graph of the previous generation of D-Wave QPUs, which has degree $3$ and no odd-length loops.
     The features of the induced Pegasus logical graph permit the benchmarking of significantly harder spin-glass instances than was possible on Chimera. 
    The induced logical graphs we examine have side length $L\in[5, 15]$, corresponding to a problem size range of $N\in[142, 1322]$ logical, error-corrected qubits.    
    
Problems on the logical QAC graph can also be encoded directly by setting $J_p=0$, 
resulting in three uncoupled and unprotected, parallel classical copies of the problem instance. 
We then extract the energies of all three copies as independent annealing samples
and denote this \q{unprotected} QA method by U3. We use the U3 method below to test whether QAC has a genuine advantage over simple classical repetition coding.\\

\noindent\textbf{\large Spin-glass instances}\\
We generate random native spin-glass instances on the induced logical graph. These types of instances have been widely used to benchmark the previous D-Wave QPUs (with the Chimera hardware graph) against classical algorithms~\cite{boixoEvidence14,speedup,katzgraberSeeking15}.
     We tested three types of spin-glass disorder: binomial, where $J_{ij}$ was randomly selected as $\pm 1$ with equal probability, Sidon-28 (S28)~\cite{katzgraberSeeking15}, where $J_{ij}$ was randomly sampled from the set $\pm\{8/28,13/28,19/28,1\}$, and finally range 6 (R6) disorder, where $J_{ij}\in \pm\{1/6,\ldots, 6/6\}$. 
    In a Sidon set, the sum of any two set members gives a number that is not part of the set. Moreover, no five numbers from the S28 set sum to zero, which prevents the occurrence of \q{floppy} qubits~\cite{q-sig}
      given the bulk degree-$5$ of the Pegasus graph.
    Binomial disorder generally admits a degenerate ground state, simplifying the optimization problem.
    In contrast, the S28 disorder can yield instances with a unique ground state~\cite{zhuBestcase16}. 
    The ground states are robust to small errors in the implementation of the $J_{ij}$ values in the binomial disorder case, but this is not the case when high precision in implementing the $J_{ij}$ values is required (as for Sidon disorder). The latter case is expected to benefit more from QAC than the former~\cite{pearsonAnalog19}. From here on, we focus on the S28 case; see Methods. \\
    
 \begin{figure*}
 \includegraphics[width=0.98\textwidth]{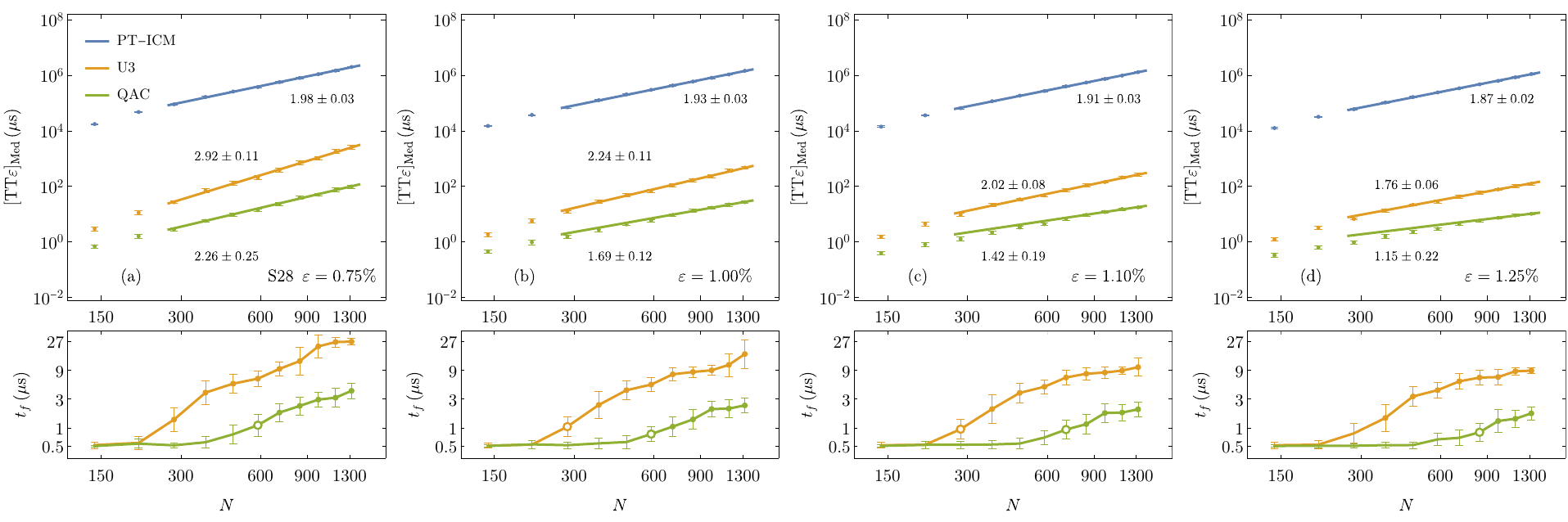}
        \caption{
            Time-to-epsilon scaling for QAC, U3, and PT-ICM for Sidon-28 (S28) spin-glass disorder.
            The bottom panels show the optimal annealing times of U3 and QAC. The top panels show the $\TTE$ results for the corresponding optimal annealing times, along with optimized PT-ICM results. The straight lines are best fits assuming power law scaling $\TTE = c N^\al$ and the accompanying numbers are the corresponding slopes $\al$. 
            As indicated in the legends, the target $\varepsilon$ increases from left to right, with a corresponding improvement in quantum annealing's performance: PT-ICM is better at low $\varepsilon$, but for $\varepsilon=1\%$, while U3 is still worse than PT-ICM, QAC already outperforms PT-ICM. 
                At $\varepsilon=1.25\%$, both QAC and U3 have better scaling than PT-ICM. 
                For higher residual energy targets, the scaling of QAC becomes unreliable since we can no longer guarantee that the optimal $t_f$ has been identified.
           We used $t_f\in[0.5, 27]\,\mathrm{\mu s}$ and $N\in[142, 1322]$. 
           Error bars for $\TTE$ data points are twice the standard error of the parameter estimate calculated using bootstrapping. 
           Filled (open) circles correspond to a $P$ value of $0.05$ ($0.20$) that $t_f^{\text{opt}}(N) > t_f^{\min}$ for the corresponding problem size $N$ (see~Methods  for details). We use only those $t_f$ values for which $P\leq 0.05$ to compute the slopes. 
        }
\label{fig:qac-results}
\end{figure*}

\noindent\textbf{\large  Time-to-epsilon metric}\\The standard performance metric for exact optimization using heuristic solvers is the time-to-solution (TTS), which quantifies the time to reach the ground state at least once with $99\%$ probability, given that the success probability per run is $p$: $\text{TTS} = t_f R$, where each annealing run lasts time $t_f$ and $R= \frac{\log(1-0.99)}{\log(1-p)}$ is the expected number of runs~\cite{speedup}. 
    To address approximate optimization, we instead consider the time required to reach 
    an energy within a fraction $\varepsilon$ of the ground state energy $E_0$, and define the time-to-epsilon for an instance as
    \begin{equation}\label{eq:tte}
        \TTE
        = t_f \frac{\log(1-0.99)}{\log(1-p_{E\leq E_0+\varepsilon \abs{E_0}})} ,
    \end{equation}
    where 
    $p_{E\leq E_0+\varepsilon \abs{E_0}}$ is the probability that the energy $E$ of a sample is no more than $\varepsilon \abs{E_0}$ above $E_0$. The TTS is the special case $\varepsilon=0$.
    In a mixed integer programming optimization context, $\varepsilon$ is known as the \emph{optimality gap}~\cite{MIPgap}, which is how we refer to it here. The ground state energies are known for our instances (see Methods ), so $\varepsilon$ is exactly calculated for each sample rather than bounded.
    An alternative time metric is the residual energy density from the ground state \cite{kingQuantum22}; we focus on the optimality gap due to its relevance to benchmarking approximate optimization algorithms.
    
    We define $[\TTE]_q$ of an instance class as the $q$-th quantile of $\TTE$ over the entire instance class. 
    Here, we focus only on the median quantile, $q=0.5$, denoted $[\TTE]_{\mathrm{Med}}$.
    For a given disorder, instance size, and $\varepsilon$-target, we find the annealing time $t_f$ (and penalty strength for QAC) that minimizes $[\TTE]_{\mathrm{Med}}$. We restrict the penalty coupling strengths to the set $J_p\in\{0.1, 0.2, 0.3\}$ to reduce resource requirements for parameter optimization, as $J_p=0.2$ is the penalty strength that most frequently optimizes the success probabilities of individual instances, and the dependence on $J_p$ above $0.2$ becomes weak.\\

\noindent\textbf{\large  Fitting the $\TTE$}\\
Below, we fit the $\TTE$ to a power law: $\TTE(N) = cN^\al$, where $\al$ is the \emph{scaling exponent}, the quantity we use to quantify the scaling of the different algorithms we compare. The choice of a power law is motivated by the existence of an $O(N)$ classical algorithm for the residual energy density; we describe such an algorithm in Methods  (though this algorithm is utterly impractical due to its huge prefactor). Due to the power law fit, we should account for factors that can modify the scaling exponent. Indeed, we could use all $N_{\max}$ qubits of the QPU and embed ${N_{\max}}/{N}$ parallel copies of each problem of size $N$, then select the best of these copies. Since, in reality, we work with only one copy due to a small fraction of the qubits and couplers being absent, we multiply the $\TTE$ by a factor of ${N}/{N_{\max}}$~\cite{speedup}. The U3 $\TTE$ is similarly multiplied by a constant factor of $3/4$ since, due to a lack of needed couplings, each instance is repeated only over the data qubits, thus leaving $1/4$ of the available (penalty) qubits unused.\\

\noindent\textbf{\large  Parallel tempering algorithm}\\
Our baseline classical algorithm is parallel tempering with isoenergetic cluster moves (PT-ICM)~\cite{zhuEfficient15}. The runtime of this algorithm has the best scaling with problem size known in the task of finding the ground state of various benchmark problems on D-Wave QPUs~\cite{mandraPitfalls17,mandraDeceptive18},
     with the only known exception being certain XORSAT instances for which highly specialized solvers have been developed~\cite{Bernaschi:2021aa,kowalsky20213regular}.
    Our optimization of the algorithmic parameters of PT-ICM is described in Methods. \\

\noindent\textbf{\large Results}\\
It is well known that the TTS metric generates unreliable results unless the annealing time is optimized for each size $N$~\cite{speedup,albashDemonstration18}.
This is because an artificially high TTS at small $N$ results in an overly flat TTS scaling. The same considerations apply to the $\TTE$, so here we find the annealing time $t_f$ that minimizes $[\TTE]_{\mathrm{Med}}$ for each $N$---denoted $t_f^{\text{opt}}(N)$---and report the resulting median $\TTE$ and its scaling estimate for QAC and U3 in \cref{fig:qac-results}, along with the analogously optimized PT-ICM results. 
    
The shortest available annealing time on the D-Wave Advantage QPU accessed via Leap is $t_f^{\min}=0.5\,\mathrm{\mu s}$, and the bottom panels of \cref{fig:qac-results} show that as the target residual energy density is increased, progressively larger problem sizes are needed to ensure that $t_f^{\text{opt}}(N)>t_f^{\min}$. 
We cannot rule out that with access to lower annealing times, one would find $t_f^{\text{opt}}(N) < t_f^{\min}$ for all $N$ values where we empirically find $t_f^{\text{opt}}(N) = t_f^{\min}$. We thus formulate a null hypothesis for each $N$ that $t_f^{\text{opt}}(N) \le t_f^{\min}$ and compute a $P$ value
as the empirical number of bootstrap samples whose $t_f^{\text{opt}}(N) = t_f^{\min}$, out of a total of $200$ samples  (see~Methods  for details of our statistical analysis). 
To compute the $\TTE$ scaling, i.e., the slope $\al$ in a fit to $\TTE = c N^\al$, for each $\varepsilon$ we use only those $t_f^{\text{opt}}(N)$ values whose $P<0.05$ (filled circles in \cref{fig:qac-results}). We can thus be confident that the reported slopes reflect the true scaling of U3 and QAC. 
    
Our first observation is that \emph{the QAC scaling is always better than the U3 scaling}, which is consistent 
with previous studies concerning the effect of analog coupling errors (\q{$J$-chaos}) on the TTS for spin-glass instances~\cite{pearsonAnalog19}. Such errors are expected for the S28 instances due to the relatively high precision their specification requires. 
 
Second, we observe that U3 and QAC reduce the absolute algorithmic runtime by four orders of magnitude compared to PT-ICM. However, this is not a scaling advantage, and since our PT-ICM calculations could be sped up by employing faster classical processors, we do not consider this a robust finding. 
Similarly, we exclude
the programming and readout time used by the D-Wave QPU. The primary source of overhead is the readout time per sample, which scales with problem size and can reach $200\,\mathrm{\mu s}$ per sample for this QPU. This timing varies by hardware generation, and its inclusion obfuscates the dominant scaling source.

Third, and most significantly, we observe that \emph{as the target optimality gap increases QA's scaling overtakes PT-ICM}. 
Notably, at $\varepsilon=1\%$, QAC exhibits a scaling exponent of $1.69 \pm 0.12$, compared to PT-ICM's $1.93\pm 0.03$, and at $\varepsilon=1.25\%$, QAC and U3 exhibit scaling exponents of $1.15\pm 0.22$ and $1.76\pm 0.06$, respectively, compared to PT-ICM's $1.87\pm 0.02$. The scaling of U3 at optimal annealing times continues to decrease as $\varepsilon$ increases; at $\varepsilon=1.5\%$ (not shown), the scaling exponents for U3 and PT-ICM become $1.60 \pm 0.07$ and $1.86 \pm 0.04$, respectively. \emph{This is robust evidence of a quantum annealing scaling advantage over the best available classical heuristic optimization algorithm}.

We are unable to determine the scaling of QAC for $\varepsilon>1.25\%$, as we cannot confirm that $t_f^{\text{opt}}(N) > t_f^{\min}$ for any $N$; as can be seen in  \cref{fig:qac-results}, already for $\varepsilon=1.25\%$ only the largest three $N$ values satisfy the $P<0.05$ criterion.
However, given the consistently better scaling of QAC for lower values of $\varepsilon$, where $t_f^{\text{opt}}(N) > t_f^{\min}$ for QAC over a range of problem sizes, it is reasonable to conclude that the QAC scaling would be a further improvement over U3 if its true $t_f^{\text{opt}}(N)$ could be established for $\varepsilon > 1.25\%$; this would require access to shorter annealing times or larger system sizes.
  
    We note that it is unsurprIsing that, given a sufficiently large target optimality gap, the D-Wave QPU returns sample energies within that gap.
    Similarly, we can expect that for large enough $\varepsilon$, even simulated annealing
    or greedy descent will be nearly guaranteed success in polynomial time.
    The significance of our result is that \emph{QA reaches near-linear scaling at a smaller optimality gap target than PT-ICM}. 
    Thus, we refer to this result as an \emph{approximate optimization advantage for quantum annealing}. 
    We also note that Ref.~\cite{kingQuantum22} similarly reported a QA optimization advantage for the residual energy density (for 3D spin glasses), but this was done at a \emph{fixed} problem size (of $N=5374$ physical qubits), and was instead concerned with the convergence of the residual energy density with the annealing time. 
    We reemphasize that, in contrast, we are reporting a scaling advantage as a function of problem size, the proper context for quantum speedup claims. 
    We explain in Methods  that our results stand regardless of whether an optimality gap or residual energy density target is chosen for benchmarking.\\
    
\noindent\textbf{\large Dynamical critical scaling}\\
As an additional perspective on the different quantum annealing dynamics resulting from error suppression, we study the difference in the dynamical critical scaling under the Kibble-Zurek (KZ) scaling ansatz~\cite{Zurek:2005aa,Campo:2018aa},
where an annealing time of $t_f \sim L^{\mu}$ is required to suppress diabatic excitations with a correlation length of $L$, and where $\mu$ is the dynamical critical exponent or KZ exponent. 
We examined signatures of dynamical critical scaling by calculating the Binder cumulant
$U = \frac{1}{2}\left(3-{\expval{q^4}}/{\expval{q^2}^2}\right)$,
where $\expval{\cdot}$ denotes the sample average, either from U3 or after decoding with QAC, and $U$ is averaged over all instances for each system size $N$ and annealing time $t_f$. Here $q = \frac{1}{N} \sum_{i=1}^{N} \sigma^z_i \sigma^{z\prime}_i$ is the overlap between two replicas, i.e., 
independently annealed $N$-spin states $\{\sigma^z_i\}$ and $\{\sigma^{z\prime}_i\}$ of a given disorder realization (set of couplings $J_{ij}$).

\begin{figure}
    \includegraphics[width=0.9\columnwidth]{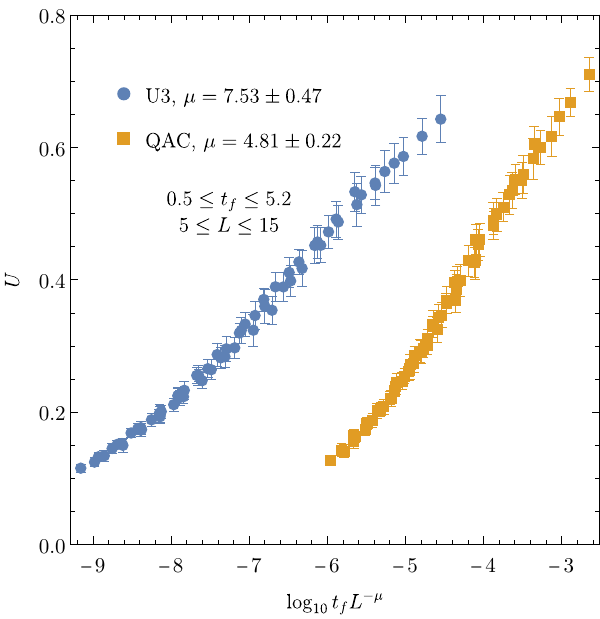}
    \caption{
        Collapsed Binder cumulant for samples collected for U3 and QAC with a penalty strength of $\lambda=0.1$. Here $t_f \in [0.5,5.2]\mu$s and $L\in[5,15]$.
}
\label{fig:binder-collapse}
\end{figure}

We performed a finite-size scaling analysis and data collapse, and the collapsed Binder cumulant for the S28 instances is shown in \cref{fig:binder-collapse}. 
We find that $\mu_{\text{QAC}} = 4.81\pm 0.22$ (at a penalty strength of $0.1$) compared to $\mu_{\text{U3}} = 7.53\pm 0.47$. The reduction of $\mu$ is lost when $\lambda=0.2$ (not shown), suggesting that $\lambda=0.1$ is optimal in the sense of diabatic error suppression. 

This effect indicates that QAC is much more effective at suppressing diabatic excitations. I.e., at equal annealing times, the dynamics are more adiabatic under QAC, in agreement with theory~\cite{Matsuura:2016aa}.
The significant reduction in $\mu$ suggests that in addition to $J$-chaos suppression, diabatic error suppression by QAC is responsible for the improved $\TTE$ and shorter optimal annealing times. Additional context is provided in Methods. \\

\noindent\textbf{\large  Discussion}\\
Using the largest available quantum annealer to date we have demonstrated an approximate optimization time-scaling advantage for quantum annealing on a family of spin-glass problems with low ground-state degeneracy and high-precision couplings. Our demonstration involves up to $1322$ logical qubits, the largest number to date in an error-corrected setting. 
Our key result is the demonstration of an \emph{algorithmic} quantum scaling advantage. 
The advantage is relative to PT-ICM, the best classical heuristic algorithm currently known for such spin glass problems, and appears at optimality gaps $\gtrsim 1\%$.

There are a few limitations to the scope of our conclusion. First, our results do not imply that finding states within small, constant gaps (and indeed finding the ground state itself) is easy for quantum annealing, nor do they imply that all spin glass problems are amenable to an approximate optimization scaling advantage via quantum annealing. Second, being finite-range and two-dimensional limits the range of applications of the problem family we have studied here. To achieve an algorithmic quantum advantage in an application setting, the next challenge for quantum optimization is demonstrating a hardware-scalable advantage in densely connected problems at sufficiently small optimality gaps.\\

\noindent\textbf{\large  Acknowledgements}\\
We thank Dr.~Evgeny Mozgunov for suggesting the use of the time-to-epsilon metric and
    Dr.~Victor Kasatkin for discussions about the theoretical lower bound of finite-range approximate optimization. 
    We also thank Dr.~Mohammad Amin, Dr.~Carleton Coffrin, Dr.~Itay Hen, and Dr.~Tameem Albash for various helpful discussions and suggestions.
    This material is based upon work supported by the Defense Advanced Research Projects Agency (DARPA) under Agreement No. HR00112190071. This research was also supported by the ARO MURI grant W911NF-22-S-0007.
    The authors acknowledge the Center for Advanced Research Computing (CARC) at the University of Southern California for computing resources.

\clearpage

\noindent\textbf{\large  Methods}\\
    
\noindent\textbf{\large Ground state energies}\\
We solved all S28 instances to optimality using Gurobi 10 within feasible runtime, except for $7$ instances of size $L=15$. 
    For these remaining instances, Gurobi proved an optimality gap of at most 1.2\%, meaning the lowest energy found is guaranteed to be no greater than $1.2\%$ above the true ground state.
    Furthermore, the lowest energies found by PT-ICM were no higher than those found by Gurobi for all instances. Thus, we assigned the ground state energies of the remaining instances to the values found by PT-ICM with high confidence.
    As we use a median over instances as our summary statistic, our conclusions are unaffected by the possibility that the ground state energy was never reached for such few instances.\\

\noindent\textbf{\large PT-ICM parameter optimization}\\
We first ran every replica once for  $N_{\text{sw}}^{\max} = 500,000$ sweeps to ensure the ground state energy was reached.
    For the largest size $L=15$, we determined $N_{\text{sw}}^{(90\%)}$, the number of sweeps that were required for $90\%$ of the instances to reach their lowest recorded energy,
    where we found  $N_{\text{sw}}^{(90\%)} \approx 31,000$ for the S28 instances.
    As this was significantly less than $N_{\text{sw}}^{\max}$,
    we considered the ground states for these instances as \q{validated} by PT-ICM to calculate median quantities over the instances.
    We finally ran PT-ICM 100 times for each instance, setting $N_{\text{sw}}=N_{\text{sw}}^{(90\%)}$.
    This yields an empirical cumulative density function for $p_{E\leq E_0+\varepsilon}$ as a function of the runtime of PT-ICM.
    The $\TTE$ is then evaluated for each instance by optimizing over the runtime of the PT-ICM repetition (where $t_f$ is now the time needed to reach the target rather than the annealing time).    
    
\begin{figure}
        \centering
        \includegraphics[width=0.95\linewidth]{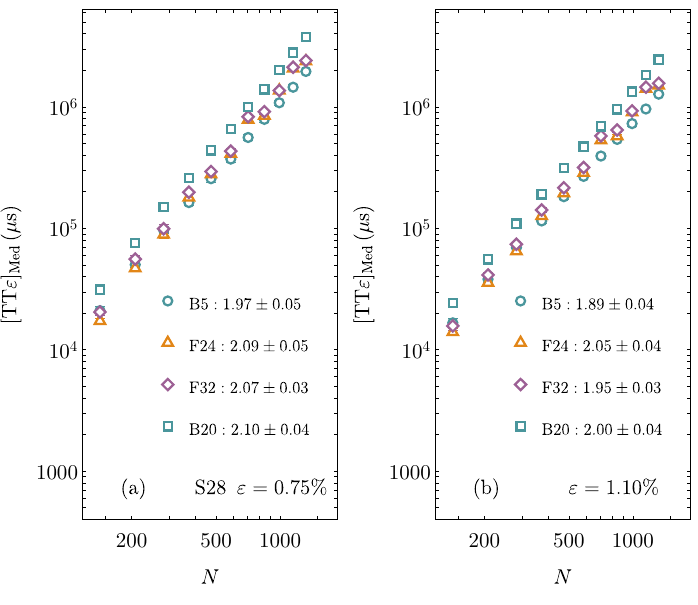}
        \caption{Time-to-epsilon of the different temperature sets used for PT-ICM and their individual scaling for $\varepsilon=0.75\%$ (left) and $\varepsilon=1.1\%$ (right). The four sets correspond, in order, to the sets shown in \cref{tab:s28-pt-temps}.
        B5: Logarithmically-spaced with $\beta_{\mathrm{max}}=5$. F24: Feedback-optimized with 24 temperatures. F32: Feedback-optimized with 32 temperatures. B20: Logarithmically-spaced with $\beta_{\mathrm{max}}=20$.
}
    \label{fig:pt}
    \end{figure}

    The scaling of PT-ICM is ideally evaluated using the parameters that best optimize the $\TTE$ for each disorder realization, instance size, and target $\varepsilon$.
    However, a rigorous optimization of the number and choice of replica temperatures for all target $\varepsilon$'s and system sizes is computationally infeasible.
    To ensure our results hold for any choice of reasonably optimized temperatures, we repeated the $\TTE$ evaluation with four temperature sets summarized in Table \ref{tab:s28-pt-temps}, which includes both logarithmically-spaced and feedback-optimized~\cite{katzgraberFeedbackoptimized06} temperatures.
    The $\TTE$ for a given disorder, instance size, and energy target was chosen from the best $\TTE$ out of the four temperature sets, illustrated in \cref{fig:pt}.
    At the optimality gap targets of interest, most temperature sets' differences are not appreciable and are unlikely to affect our conclusions.
    A more comprehensive but computationally expensive optimization of PT-ICM would involve a grid search over a range of the parameters $\beta_{\mathrm{min}}$, $\beta_{\mathrm{max}}$, and $N_{\mathrm{icm}}$.\\
    
    \begin{table}
        \begin{tabular}{c|c|ccc}
            \multicolumn{2}{c}{ }& \multicolumn{3}{|c}{\noindent\textbf{S28}}  \\
            \hline
            Set &  $N_T$  & $\beta_{\mathrm{min}}$ & $\beta_{\max}$ & $N_{\mathrm{icm}}$   \\
            \hline
            1 &32 & 0.1 & 5.0 & 8  \\
            2 &24 & 0.2 & 10.0 & 6 \\
            3 &32 & 0.2 & 10.0 & 8 \\
            4 & 32 & 0.1 & 20.0 & 8 \\
        \end{tabular}
        \caption{ 
            Temperature sets used for PT-ICM for S28 instances, 
            where $\beta_{\mathrm{min}}$ is the hottest temperature, $\beta_{\max}$ is the coldest temperature, and $N_{\mathrm{icm}}$ is the number of low temperature (largest $\beta$) subject to ICM moves. 
            The temperatures are logarithmically spaced in sets 1 and 4.
            The temperatures in sets 2 and 3 were feedback-optimized with initially logarithmically spaced temperatures.}
            \label{tab:s28-pt-temps}
    \end{table}

\noindent\textbf{\large Software Implementation}\\
Our PT-ICM implementation is available as the TAMC software package~\cite{TAMC}.
    Our implementation of QAC for the D-Wave Advantage graph is available as part of the PegasusTools Python package~\cite{PegasusTools}.

    We also implemented and considered the performance of simulated annealing but do not present its results here as this algorithm was not competitive at large problem sizes. 
        Our PT-ICM implementation is a general-purpose solver written in the Rust programming language, targeting CPU execution, and accepts instances with any connectivity.
    This is in contrast to previous studies that have used simulated annealing or the Hamze-de Freitas-Selby algorithm solvers that are specialized for problems defined on the Chimera graph~\cite{hamzeFields04,selbyEfficient14}.\\

\noindent\textbf{\large Statistical methods}\\
For a given problem size and target optimality gap, we calculate the median $\TTE$ from the quantum annealing data using a three-step Bayesian bootstrap procedure over the levels of readout samples, gauges (spin-reversal transformations at the hardware level), and instances: (1) the success probability for each gauge is resampled from a beta distribution for $N_{\text{samp}}$ samples per gauge, (2) the statistical weight of each gauge is sampled from a Dirichlet distribution of length $N_{\text{gauge}}$ to take the weighted average success probability for each instance, and (3) the statistical weight of each instance is sampled from a Dirichlet distribution of length $N_{\text{inst}}$ to take the weighted median.
We performed our quantum annealing experiments with $N_{\text{samp
}}=1000 \mathrm{(QAC)}/3000 \mathrm{(U3)}$, $N_{\text{gauge}}=10$, and $N_{\text{inst}}=125$. We performed $N_{\text{boots}}=200$ bootstrap samples per size and energy density target pair and found the annealing time and penalty strength that resulted in an optimal $\TTE$ for each bootstrap sample.
The distribution of the optimal median $\TTE$ values and the distribution of optimal annealing parameters are the two final products of this sampling procedure shown in \cref{fig:qac-results}, with $2\sigma$ error intervals for the optimal $\TTE$.

Next, we formulate a null hypothesis and compute $P$ values as follows. 
    The null hypothesis is that the optimal annealing time $t_f$ is the minimum accessible $t_f^{\min} = 0.5\,\mathrm{\mu s}$, i.e., that the true optimal annealing time is not above $t_f^{\min}$.
    The $P$ value is the empirical number of bootstrap samples whose optimal $t_f$ was $0.5\,\mathrm{\mu s}$, out of 200 samples.
Filled circles in \cref{fig:qac-results} mean $P<0.05$ for the probability that $t_f=0.5\mu s$ in the bootstrap sample, while open circles mean $P<0.20$. The filled circles show which points have the highest confidence that the optimal annealing time is not below $0.5\,\mathrm{\mu s}$.\\

\noindent\textbf{\large Non-universality of the observed KZ critical exponent}\\
The Binder cumulant $U$ is well-known to provide a statistical signature of phase transitions. Under the dynamic finite-size scaling 
ansatz, $U(L, t_f)$ is expected to collapse onto a common curve for all system sizes $N$
when $t_f$ is rescaled by $L^{-z-1/\nu}$, where $\nu$ and $z$ are the correlation length and dynamic critical exponents, respectively. This reflects the KZ ansatz: the annealing time required for the system to remain adiabatic up to a correlation length of $L$ scales as
\beq
t_f(L) \sim L^{\mu}, \quad \mu = z + \frac{1}{\nu} .
\eeq

While the collapse seen in \cref{fig:binder-collapse} is visually convincing, the error estimates for $U$ are, unfortunately, too large to determine the KZ exponent or to extract $\nu$ and $z$. 
In addition, we do not observe that the extracted KZ exponent is universal: the estimate for $\mu$ that best collapses the Binder cumulant for binomial (as opposed to S28) spin glass instances is $\mu\approx 9\pm1$ for U3, and is reduced to $\mu \approx 7.5\pm 0.6$ with QAC at $\lambda = 0.1$. Thus, while the scaling ansatz yields a useful complementary way to quantify the advantage of QAC in terms of sampled quantities in addition to the $\TTE$ metric, the lack of universality and the possibility that the spin-glass transition temperature is zero~\cite{katzgraberGlassy14} do not clearly support a universal critical description of the annealing dynamics, quite unlike the conclusions of Refs.~\cite{kingCoherent22-2,kingQuantum22-2}.\\

\begin{figure}
    \centering
    \includegraphics[width=0.8\linewidth]{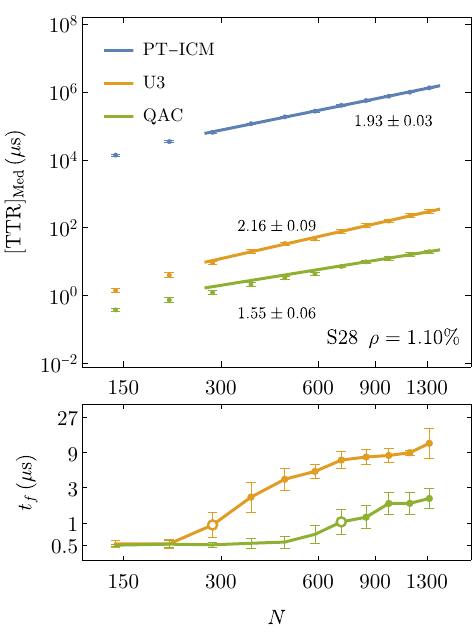}
    \caption{\label{fig:ttr}Time-to-rho performance and scaling of S28 instances, which are comparable to those using the time-to-epsilon metric.}
\end{figure}

\noindent\textbf{\large Time-to-residual energy}\\
Ref.~\cite{kingQuantum22} performed energy decay measurements of QA dynamics in 3D spin glass instances as captured through the residual energy, a dimensionless quantity, 
\begin{equation}
    \rho = \frac{\expval{H_z} - E_0}{JN} ,
\end{equation}
where $N$ is the number of spins and $J$ is the characteristic energy scale of the Ising Hamiltonian (which is simply 1 for all of our instance classes). This motivates an alternative measure for approximate optimization, which we call \emph{time-to-residual energy}, or \emph{time-to-rho},
 \begin{equation}
    \mathrm{TTR}(N)= t_f \frac{\log(1-0.99)}{\log(1-p_{E \leq E_0 + \rho J N})}\,,
\end{equation}
where now $\rho$ sets the target energy difference from the ground state in units of $JN$, rather than $E_0$ as in the case of $\TTE$. Nevertheless, both $\varepsilon$ and $\rho$ are targets that grow in proportion to the problem size or number of variables (in finite dimensions). The TTR analog of \cref{fig:qac-results} at $\rho=1.1\%$ is shown in \cref{fig:ttr}. 
While the precise scaling exponents $\al$ of U3 and QAC vary slightly, the optimal annealing times are nearly identical. The trends of the scaling exponents as functions of either $\rho$ or $\varepsilon$ are also qualitatively similar. 
In finite-dimensional spin glasses, the variance of $E_0$ across instances appears to scale with $\sqrt{N}$ in the thermodynamic limit \cite{bouchaudEnergy03}. Hence, $E_0/N J$ converges to a constant, and there is effectively no distinction between $\varepsilon$ and $\rho$ in the thermodynamic limit.\\

\noindent\textbf{\large Classical complexity of approximate optimization}\\
Under the TTR metric, it can be shown that the approximate optimization of finite-range spin glasses has, in theory, a linear scaling using a simple divide-and-conquer algorithm. Namely, for a given residual energy target $\rho$, an algorithm exists whose TTR scales linearly in the system size for a sufficiently large size, with a large prefactor depending on the residual energy. For 2D spin glasses on a square lattice, it can be summarized as follows:
\begin{enumerate}
\item Partition the size-$N=L^2$ instance into $K\times K$ subgraphs $G_{x,y}$, with $x,y\in\{1,\ldots,K\}$, each with \emph{constant} side length of $L_0 = L/K$ spins;
\item Find the local ground states 
for each subgraph using an exact or heuristic solver 
for each $L_0\times L_0$ instance. This step has cost $C(L_0)$, potentially exponential in $L_0^2$ on a non-planar lattice;
\item Patch together all subgraph ground state configurations as the approximate ground state for the global Hamiltonian and return this state's energy as the approximate ground state energy $E^*(N)$. This last step requires $O(N)$ time due to the need to sum the energies of $K^2 = O(N)$ patches.
\end{enumerate}
The overall complexity is, therefore, $C(L_0) O(N)$.

This algorithm optimizes \emph{bulk} energies throughout the volume of the spin-glass at the expense of large energy violations over regions scaling with the surface area of the spin-glass, i.e., at the patch boundaries. The boundary excitations become negligible for sufficiently large sizes compared to the bulk energies, with the latter eventually reaching the desired residual energy.
More precisely, up to $4L_0$ boundary couplings may be violated per volume of $L_0^2$, so the residual energy density for this algorithm is upper-bounded by $4/L_0$. 
Thus, we estimate that the regime of system sizes where this algorithm applies for the $\rho$ targets we examine, e.g., 1\%, would require a reliable, efficient solver for instances with at least a \emph{sub-problem} side length of $L_0 \approx 400$, resulting in a prefactor of $2^{400}\sim 10^{120}$, which is entirely impractical.
Nevertheless, this algorithm could be a starting point for parallel and quantum-classical hybrid algorithms for massive, finite-dimensional problems.
Furthermore, such an algorithm could reach a similar target $\varepsilon$ in linear time with increasing probability as the system size increases due to the decreasing variance of $E_0/(JN)$.\\

\begin{figure}
    \centering
    \includegraphics[width=\linewidth]{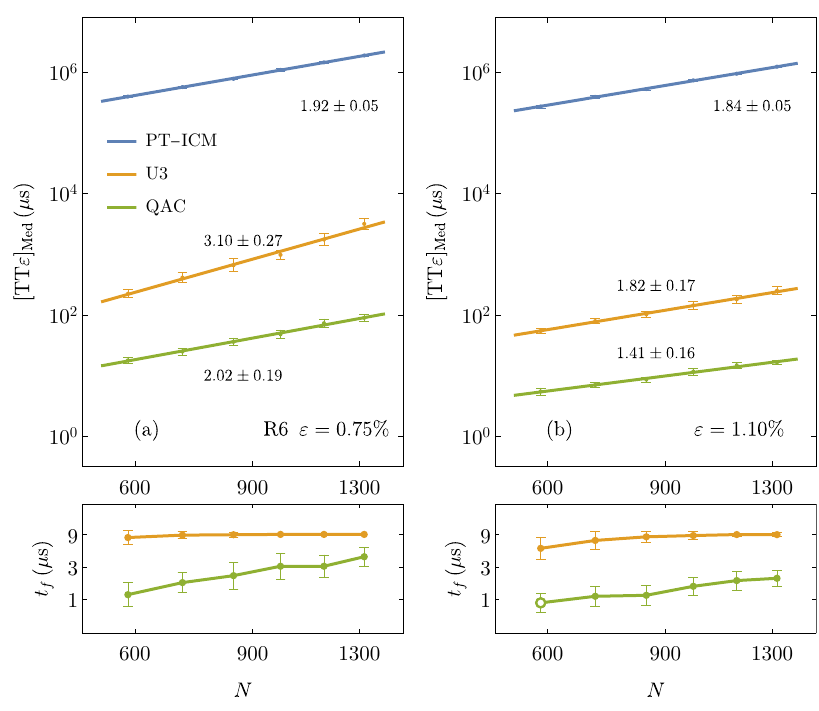}
    \caption{\label{fig:u6}
        Time-to-epsilon scaling of spin glass instances with R6 disorder for (a) $\varepsilon=0.75\%$ and (b) $\varepsilon=1.10\%$.
        To expedite finding the optimal $\TTE$,
        the annealing times were limited to the range $t_f\in [0.5, 9]\,\mathrm{\mu s}$ and QAC was limited to $J_p=0.2$ only.}
\end{figure}

\noindent\textbf{\large Alternative spin-glass disorder cases}\\
We performed a similar study of instances with binomial disorder $J=\pm 1$. UnsurprIsingly, there was no substantial difference in scaling between U3 and QAC, and additionally, we were unable to validate the optimal annealing time for QAC at $\varepsilon=1\%$. 
Thus, we considered the binomial disorder instances unsuitably easy for our purposes.

In an attempt to estimate the influence of precision and ground state degeneracy in the S28 disorder, we also studied range 6 (R6) instances for which the couplings were randomly drawn from $J\in \{\pm1/6, \pm2/6, \pm3/6, \pm4/6, \pm5/6, \pm1\}$. 
The main contrast between R6 and S28 is that the minimum non-zero local field a qubit may experience is much smaller in S28 disorder. Furthermore, in the absence of the Sidon property, the R6 disorder is more susceptible to ground-state degeneracies due to floppy qubits. However, it is less likely to occur than in a smaller range disorder case such as range 3.

The $\TTE$ scaling for the R6 case is shown in \cref{fig:u6}. 
For a smaller $\varepsilon=0.75\%$, the scaling of QAC is better with R6 disorder than S28 disorder (\cref{fig:qac-results}), going from 2.26 to 2.02, then equalizes as $\varepsilon$ increases.
Perhaps surprIsingly, the opposite is true of U3, which scales worse on R6 disorder at smaller epsilon (2.92 to 3.10) before scaling better than S28 disorder above $\varepsilon \approx 1.1\%$ (2.02 to 1.82).
That is, while S28 is overall more challenging than R6 under QAC, low energy sampling of R6 is more challenging for unprotected QA than S28. This is likely caused by the worse relative precision of small coupling strengths in unprotected QA, which would not affect the S28 disorder.\\

\end{document}